\documentclass[%
 reprint,
 amsmath,amssymb,
 aps,
pra,
]{revtex4-1}

\usepackage[normalem]{ulem}%subrayar, tachar, etc
\usepackage{graphicx}% Include figure files
\usepackage{dcolumn}% Align table columns on decimal point
\usepackage{bm}% bold math
\usepackage{color}
\begin{document}

%\preprint{APS/123-QED}

\title{Controlled generation of mixed spatial qudits with arbitrary degree of purity}% Force line breaks with \\

\author{J.~J.~M.~Varga}
 \email{miguel@df.uba.ar}
\author{S.~Ledesma}%
\author{C.~Iemmi}%
\affiliation{Universidad de Buenos Aires, Facultad de Ciencias Exactas y Naturales, Departamento de F\'isica, Bue­nos Aires, Argentina.}
\affiliation{Consejo Nacional de Investigaciones Cient\'ificas y T\'ecnicas, Bue­nos Aires, Argentina.}
%\\ This line break forced with \textbackslash\textbackslash}%


\author{L.~Reb\'on}
\affiliation{
 Departamento de F\'isica, IFLP, Universidad Nacional de La Plata, C.C. 67, 1900 La Plata, Argentina.}%
%\affiliation{
 %Institut f\"ur Laserphysik,
%Universit\"at Hamburg, Luruper Chaussee 149, 22761 Hamburg, Germany.
%\affiliation{% Authors' institution and/or address\\


\date{\today}% It is always \today, today,

\begin{abstract}
We propose %and experimentally show
a method for preparing mixed
quantum states of arbitrary dimension $D$ ($D\geq2$) which are
codified in the discretized transverse momentum and position of
single photons, once they are sent through an aperture with $D$
slits. Following our previous technique we use a programmable single
phase-only spatial light modulator (SLM) to define the aperture and
set the complex transmission amplitude of each slit, allowing the independent
control of the complex coefficients that define the quantum state.
Since these %programable devices 
SLMs give us the possibility to dynamically varying the
complex coefficients of the state during the measurement time, we can generate 
not only pure states but also quantum states compatible with a mixture of pure quantum states. 
Therefore, by using these apertures %(against static ones) 
varying on time according to a probability distribution, we have experimentally obtained $D$-dimensional quantum states with purities that depend on the parameters of the distribution
through a clear analytical expression. This fact allows us to easily customize the
states to be generated. Moreover, the method offer the possibility of working without changing the optical setup between pure and mixed states, or when the dimensionality of the states is increased. The obtained results show a quite good performance of our method at least up to
dimension $D=11$, being the fidelity of the prepared states $F > 0.98$ in every case.   

\end{abstract}
%Moreover, the method offer the possibility of working without change the optical setup and without significantly increasing the consumed experimental time when the dimensionality of the states is increased. The obtained results show a quite good performance of our method at least up to
%dimension $D=11$, being the fidelity of the prepared states $F > 0.98$ in every case.   

%With this technique we have achieved 
%
%With the same optical setup and without significantly
%increasing the consumed experimental time, we have obtained quite
%good performances of our technique for several states up to
%dimension $D=11$, being the fidelity of the prepared states $F >
%0.97$ {\color{red}[Ver cu\'al es realmente este valor]} in each case.

\maketitle

%\tableofcontents

\section{\label{sec:intro}Introduction}

%-----------------------------------------------
In quantum optics, pure quantum states of single photons have been widely explored, both theoretically and experimentally. They can be generated, controlled and measured using the several degrees of freedom of a photon, and by means of different techniques \cite{Kwiat1995,Kwiat1999,kokBook}.
However, a quantum system is not in general in a pure state. Because of experimental imperfections
or interactions with the environment, we have only partial knowledge of its physical state and it cannot be described through a well defined vector $|\psi\rangle$ in the Hilbert space. For that reason, the
most general description of a quantum system is given by a \textit{mixture} of pure quantum states that can be matematicaly expressed by the formalism of the density matrix \cite{Fano1957}. In consequence, the progress in the study of quantum systems and their potentialities for
practical applications, relies on the ability for controlling mixed states, and not only pure states. For instance, the ability for engineering and measuring mixed quantum states allows to experimentally study how
quantum computing algorithms and quantum communication protocols 
are affected by decoherence \cite{zurek02,kendon2007decoherence}. Besides, beyond the original model for 
quantum information \cite{NielsengBook,jaeger2007Book}, 
based in unitary gates operating on pure quantum states, alternative models based on mixed quantum states have been developed\cite{Laflamme1998,Datta2008}. These models also give the possibility to perform some tasks not realizable with a comparable classical system \cite{Meyer2000,Roa2011,dakic2012,frey2013}.
%with a computing speed-up over the best classical algorithm {\color{red}[Ref]}. 
% It is not known, however, if all informationprocessing tasks that can be done more efﬁciently with a quantum system than with a comparable classical system require entanglement as a resource. 
Moreover, as the system is initially in a mixed quantum state, and entanglement is not the required physical resource, 
they are less restrictive, more robust against noise, and easier to implement than the standard quantum information model.

Controllable generation of mixed quantum states has been  successfully proposed in earlier works, mainly, using the polarization degree of freedom to codified the state \cite{baek2011preparation,lanyon2008experimental,Englert2013mix,Rebon2016} . 
%Different methods were used to control the level of mixture. In \cite{baek2011preparation,lanyon2008experimental}, the initial (pure) qubit is sent through a polarization interferometer. An arbitrary mixed state is prepared as an incoherent, or partially coherent superposition of the two output modes by changing the path-length difference between the arms of the interferometer. Alternatively, mixed qubit states can be obtained by using variable polarization rotators \cite{Englert2013mix,Rebon2016} {\color{red}[¿Dejo nuestra Ref?]}. In such scheme, the purity is controlled by switching the rotators between two positions, and averaging over a sufficient interval of time to sample both polarization states. {\color{red}[Incluir este párrafo si se quiere explicar los métodos]} 
While these methods are relatively simple to implement, they only allow the realization of two-level systems. Otherwise, higher dimensional quantum states, namely qudits ($D$-level quantum systems), increase the quantum complexity without increasing the number of particles involved. For instance, systems of dimension $D = 2^N$ can be used to simulate a composite system of $N$ qubits \cite{lima2011}. For quantum communication protocols, $D$-dimensional quantum channels show higher capacity, and provide better security againts an eavesdropper \cite{Bechmann2000,Cerf2002,Wang2005}. Moreover,  multi-level information carriers are crucial to reduce the number of gates required in the circuits for quantum computing \cite{lanyon2009simplifying}. 
 
Among the feasible degrees of freedom for encoding  high-dimensional quantum systems \cite{Mantaloni2009,Torner2007,Neves2004,
Boyd2005}, the discretized transverse momentum-position of single photons have attracted particular interest. They have proven useful for several application such as quantum information protocols \cite{Solis2011,Solis2017}, quantum games \cite{Kolenderski2012}, quantum algorithms \cite{Marquez2012}, and quantum key distribution \cite{etcheverry2013}. The encoding process is achieved by sending the photons through an aperture with $D$ slits, which sets the qudits dimension \cite{Neves2005}. More sophisticated methods to generate these so-called \textit{spatial qudits}, take advantage of liquid crystal displays (LCDs) as programmable spatial light modulators (SLMs). These programmable optical devices can be used to define a set of independent $D$ slits with complex transmission. In this way, it is possible to produce
and measure \textit{arbitrary pure qudits} without any extra physical alignment of the optical components~ \cite{lima2009manipulating,lima2011,solis2013,
varga2014}. 

Recently, Lemos \textit{et al} \cite{Lemos2014} have characterized
the action of an SLM as a noisy quantum channel
acting on a polarization qubit, and they used it for implementing a phase flip channel with a controllable degree of decoherence. In Ref.~\cite{Padua2015} Marques \textit{et al} extended the use of the SLMs to simulate the open dynamics of a $D$-dimensional quantum system by using films instead of images. 

In this paper, we present a method to generate \textit{arbitrary spatial mixed states} of $D$ dimension $(D \geq 2)$, which is based in the techniques developed in our previous works~\cite{solis2013,varga2014}. 
%It follows the proposals in Refs.~\cite{Moreno99,Bagnoud2004} for encoding amplitude and phase information onto a phase-only SLM. It allow us to control independently the complex coefficients which define the quantum state. To this end, phase diffraction gratings were displayed in those zones corresponding to the slits for setting the amplitudes of the complex transmission. The imaginary part is fixed by adding a \textit{constant phase} value to the phase grating. After a spatial filtering process, a \textit{pure} qudit state is obtained in the image plane. 
We have extended these techniques to consider a $D$ slits with a variable complex transmission. 
The use of a programmable SLM makes it possible to \textit{dynamically modify} the complex transmission, in order to obtain a mixed qudit state by averaging the sample over the time.
  
The paper is organized as follows: In Section \ref{sec:formalism} we give the mathematical description of a single photon state when it is sent trough an aperture with a time-varying transmission function. By considering that we can vary the relative phase values of the complex transmission following an uniform probability distribution, we have derived simple analytical expressions, which show the dependence between the distribution widths and the purity of the state, for any dimension $D$. From these expressions it is possible to obtain any degree of purity by continuously varying the the distribution widths, which allows us to use the same method for preparing pure and mixed states. 
In Section \ref{sec:exp} it is described the experimental set-up and it is explained how a first SLM is addressed to generate the states, while a second SLM is employed to encode the measurement bases used to perform the tomographic reconstruction of the system. In Section \ref{sec:pure} it is reported a first experiment, consisting in the generation and measurement of pure qudit states. It was carried on in order to test the set-up and the proposed methods.
%In Section \ref{sec:exp} we start with the description of the experimental set-up and how the SLM is used to generate the states. We also describe how a second SLM is used to encode the measurement bases employed to perform the tomographic reconstruction of such systems as described in Ref. \cite{lima2011}. We first test, in Section \ref{sec:pure}, the generation and measurement stages of our set-up by preparing pure qudit states.%
Afterwards, in Section \ref{sec:mixed}, we implement the variable transmission function for generating mixed states with different degree of purity in dimensions $D$=2, 3, 7 and 11. Finally, the results are presented in Section \ref{sec:results_mix} and discussed before go into the conclusions.

%\cite{2008LimaMix}

%-----------------------------------------------
\section{\label{sec:formalism}Formalism}

Let us to start by considering the generation of a spatial qudit in a pure state. A paraxial and monochromatic single-photon field is transmitted
through an aperture described by a complex transmission function
$A(\mathbf{x})$. Assuming an initial pure state, $|\psi\rangle$, it is transformed as

\begin{eqnarray} \label{eq:photon_transmitted}
|\psi\rangle=\int\!d\mathbf{x}\,\psi(\mathbf{x})|1\mathbf{x}\rangle
\;\;\stackrel{A(\mathbf{x})}{\Longrightarrow}
\;\;\int\!d\mathbf{x}\,\psi(\mathbf{x})A(\mathbf{x})|1\mathbf{x}\rangle,
\end{eqnarray}
where $\mathbf{x}=(x,y)$ is the transverse position coordinate and
$\psi(\mathbf{x})$ is the normalized transverse probability
amplitude for this state, i.e.,
$\int\!d\mathbf{x}\,|\psi(\mathbf{x})|^2=1$.

We are interested in generating an incoherent mixture of pure states by varying the transmission function of the aperture over time. So, let us consider that
$A(\mathbf{x},t)$ is an array of $D\geq 2$ rectangular slits of
width $2a$, period $d$ and length $L(\gg a,d)$, where each slit,
$\ell$, has a transmission amplitude $\beta_\ell(t)$:

\begin{eqnarray}    \label{eq:Dslits}
A(\mathbf{x})\rightarrow A(\mathbf{x},t)&=& {\rm
rect}\!\left(\frac{x}{L}\right)\times\sum_{\ell=0}^{D-1}\beta_\ell(t)\;{\rm
rect}\!\left(\frac{y-\eta_\ell d}{2a}\right),\nonumber\\
\end{eqnarray}
with $\eta_\ell=\ell+(D-1)/2$.

Thus, instantaneously, at any time $t$ a pure state $|\psi(t)\rangle$ is obtained, whereas in a finite
period of time $\Delta t$, an ensemble of these pure states is created. In consequence, the result after a measurement process is the ensemble average over the integration time $T$, whose statistics corresponds to a mixed state described by the density matrix \cite{Fano1957}, $\rho$:

\begin{eqnarray} \label{eq:mixed}
\rho&=&\frac{1}{T}\int_0^T dt|\psi(t)\rangle\langle\psi(t)|\nonumber\\
&=&\int d\mathbf{x} \int
d\mathbf{x'}\rho(\mathbf{x},\mathbf{x}')|1\mathbf{x}\rangle\langle
1\mathbf{x}'|,
\end{eqnarray}
where
$\rho(\mathbf{x},\mathbf{x}')\equiv\psi(\mathbf{x})\psi^*(\mathbf{x}')\frac{1}{T}\int_0^T
dt A(\mathbf{x},t)A^*(\mathbf{x},t)$. Hence, the state of the photon
in Eq.~(\ref{eq:mixed}) can be written as

\begin{eqnarray}
\label{eq:spatial_qudit}
\rho=\sum_{\ell,\ell'=0}^{D-1}\tilde{c}_{\ell,\ell'}|\ell\rangle\langle\ell'|,
\end{eqnarray}
where $|\ell\rangle$ denotes the state of the photon passing through the slit $\ell$ \citep{Neves2004}.
The states $|\ell\rangle$ satisfy the condition
$\langle\ell|\ell'\rangle=\delta_{ll'}$, and they are used to define the logical base for spatial qudits. The quantum state of the system is determined by the coefficients  $\tilde{c}_{\ell,\ell'}=\frac{1}{T}\int_0^T dt ~ \beta_{\ell}(t)\beta^*_{\ell'}(t)$, which carry the information codify in the transfer function $A(\mathbf{x},t)$. In principle, given that in the most general case the transmission amplitudes $\beta_{\ell}(t)$ are complex values, we could introduce the time dependence either in the modulus, $\vert\beta_{\ell}(t)\vert$, or in the argument, Arg$\left( \beta_{\ell}(t)\right)$, and even in both. However, as it is well known, phase information plays a more important role than real amplitude in signal processing \cite{oppenheim1981} so we can get full control of the state by varying only the phases (see Sec. \ref{sec:mixed} for a complete discussion). Then, for a time-dependent phase, $\phi_\ell(t)$, the transmission for the slit $\ell$ is written as $\beta_\ell(t)=\beta_\ell~e^{i\phi_\ell(t)}$, and the complex coefficients in the mixture in Eq. (\ref{eq:spatial_qudit}) are given by the expression 

\begin{eqnarray}\label{eq:complex_coeff}
\tilde{c}_{\ell,\ell'}&=&\left(\beta_\ell\beta_{\ell'}/\sqrt{\sum_{j=0}^{D-1}\beta_j^2}\right)\times~c_{\ell,\ell'},
\end{eqnarray}
with
\begin{eqnarray}\label{eq:c_time}
c_{\ell,\ell'}&=&\frac{1}{T}\int_0^T dt ~ e^{~i\phi_{\ell}(t)}
e^{-i\phi_{\ell'}(t)}.
\end{eqnarray}

To define this state we have proposed that the phase of each slit varies according to a probabilistic distribution.  If the time $T$ is much longer than the characteristic time where the phase varies, the integration in the time domain can be replaced by an integration in the phase domain, $\Omega$. In fact,
we can assume that for a period of time long enough, $\phi_{\ell}(t)$ reaches all its possible values
with a frequency of occurrence given by a probability distribution $f(\alpha_{\ell})$ $(\alpha_{\ell}\in\Omega)$. In
addition, if the phase of each slit varies independently of the other ones, the
joint probability distribution is obtained as

\begin{eqnarray}\label{eq:joint_distr}
f(\mathbf{\alpha})\equiv
f(\alpha_0,\alpha_1,...,\alpha_{D-1})=f(\alpha_0)f(\alpha_1)...f(\alpha_{D-1}).\nonumber\\
\end{eqnarray}
According to this scheme, the complex coefficients in Eq. (\ref{eq:c_time})
turn into

\begin{equation}\label{eq:coefficients}
c_{\ell,\ell'}=\int d\mathbf{\alpha} ~ f(\mathbf{\alpha})
e^{~i\alpha_{\ell}} e^{-i\alpha_{\ell'}}.
\end{equation}
From this expression we directly obtain
$c_{\ell,\ell}=1,~\forall \ell=0,1...,D-1$ and $c_{\ell,\ell'}=c_{\ell',\ell}^*,~\forall \ell,\ell'=0,1...,D-1$, implying that the diagonal elements of the density matrix (Eq. (\ref{eq:spatial_qudit})), which denote the probabilities to find the system in one of the
(pure) quantum states $|\ell\rangle$, are real-valued coefficients in the interval $\left[0,1\right]$, and as expected, the density matrix is Hermitian ($\rho^{\dag}=\rho$).

As only the relative phases (but not the absolute values) in the linear combination
that define the quantum state are relevant,
we have (indistinctly) fixed the phase value of one of the slit, $|0\rangle$, to be $\phi_0=0$.
Then, the corresponding probability distribution in Eq. (\ref{eq:joint_distr}) is the
Dirac delta function $\delta(\alpha_0)$. Besides, we have assumed that
$f(\alpha_{\ell})$ is a uniform distribution of width $\Delta_{\ell}$
and centered in $\alpha_{\ell}=\phi_{\ell}, ~\forall \ell=1...,D-1$.
In this way, the joint probability distribution is

\begin{eqnarray}
f(\mathbf{\alpha})&=&\mathcal{N}\times\delta(\alpha_0)\prod_{{\ell}=1}^{D-1}{\rm
rect}\left(\frac{\alpha_{\ell}-\phi_{\ell}}{\Delta_{\ell}}\right),\\
\mathcal{N}&=&\frac{1}{\prod_{j=1}^{D-1}\Delta_j}.\nonumber
\end{eqnarray}

Therefore, the statistical mixture $\rho$ which describe the state of the transmitted photon, will be completely
determined by the real amplitudes $\beta_{\ell}$, the phases $\phi_{\ell}$, and the distribution widths $\Delta_{\ell}$, which can
be completely and independently controlled in our experimental setup (see Sec.~\ref{sec:exp}).
Even more, it is straightforward to obtain the purity of the state, $P\left(\rho\right)\equiv\mathrm{Tr}\left(\rho^2\right)$:

\begin{eqnarray}\label{eq:purityD}
P\left(\rho\right)&=&\mathcal{Z}^2~\sum_{i=0}^{D-1}\beta_i^4\nonumber\\
&+&2~\mathcal{Z}^2\sum_{i=0}^{D-2}\prod_{j>i}^{D-1}\beta_i^2~\beta_j^2~\mathrm{sinc}\left(\frac{\Delta_i}{2}\right)^2
\mathrm{sinc}\left(\frac{\Delta_j}{2}\right)^2,\nonumber\\
\end{eqnarray}
being the normalization constant $\mathcal{Z}=\left(\sum_{i=0}^{D-1}\beta_i^2\right)^{-1}$.
From this equation (Eq.~(\ref{eq:purityD})) it becomes clear how to generate a qudit state with an arbitrary purity, by
setting up the experimental parameters. In particular, the maximal mixed state $(P\left(\rho\right)=\frac{1}{D})$ is obtained when the real coefficients $\beta_{\ell}$
have all of them the same value, and the phases can reach any value between $0$ and $2\pi$ with the same probability, i.e., $\Delta_{\ell}=2\pi, ~\forall \ell=1...,D-1$.
On the other hand, if $\Delta_{\ell}\rightarrow 0, ~\forall \ell=1...,D-1$, i.e., when the phase of each slit
remains constant over the time $T$, the terms $\mathrm{sinc}\left(\frac{\Delta_{\ell}}{2}\right)$
are equal to 1. In such a case, the purity of the state tends to 1, as expected for a pure quantum state.
Thus, the scheme discussed here is reduced to the previous ones presented in Refs.~\cite{solis2013} and \cite{varga2014} for preparing arbitrary pure spatial qudits.

Let us consider as example the preparation of qubit states. Because of their simplicity,
they are helpful to understand the general behaviour of the scheme. In this case  --$D$=2-- we explicitly obtain

\begin{eqnarray}
c_{01}&=&\frac{1}{\Delta_1}\int_{\phi_1-\frac{\Delta_1}{2}}^{\phi_1+\frac{\Delta_1}{2}}d\alpha_1~
e^{-i\alpha_1}=
e^{-i\phi_1}~\mathrm{sinc}\left(\frac{\Delta_1}{2}\right).\nonumber\\
\end{eqnarray}
The diagonal coefficients of the density matrix are independent of the phase probability distribution,
since as was mentioned before $c_{\ell,\ell}=1$, while the rest of coefficients can be obtained by complex conjugation. Therefore, these states are described by the density matrix

\begin{eqnarray}
\rho=\frac{1}{\beta_0^2+\beta_1^2}\left(%
\begin{array}{cc}
 \beta_0^2  & \beta_0\beta_1e^{-i\phi_1}\mathrm{sinc}\left(\frac{\Delta_1}{2}\right)\nonumber\\
 \beta_0\beta_1e^{i\phi_1}\mathrm{sinc}\left(\frac{\Delta_1}{2}\right) & \beta_1^2
\end{array}%
\right).
\end{eqnarray}
They have a purity given by

\begin{eqnarray}\label{eq:purity2}
P\left(\rho\right)=\frac{\beta_0^4+\beta_1^4+2\beta_0^2\beta_1^2\mathrm{sinc}^2\left(\frac{\Delta_1}{2}\right)}{(\beta_0^2+\beta_1^2)^2},
\end{eqnarray}
and any degree of purity can be achieved by controlling the relation $\beta_1/\beta_0$, and $\Delta_1$.

In the next section we described our technique developed for implementing these concepts, and illustrate with the generation of mixed states in different dimensions $D$.

%--------------------------------------------------
\section{\label{sec:impl}Experimental implementation}

\subsection{\label{sec:exp}Experimental Set-up}

\begin{figure}[htbp]
\centering
\includegraphics[width=\linewidth]{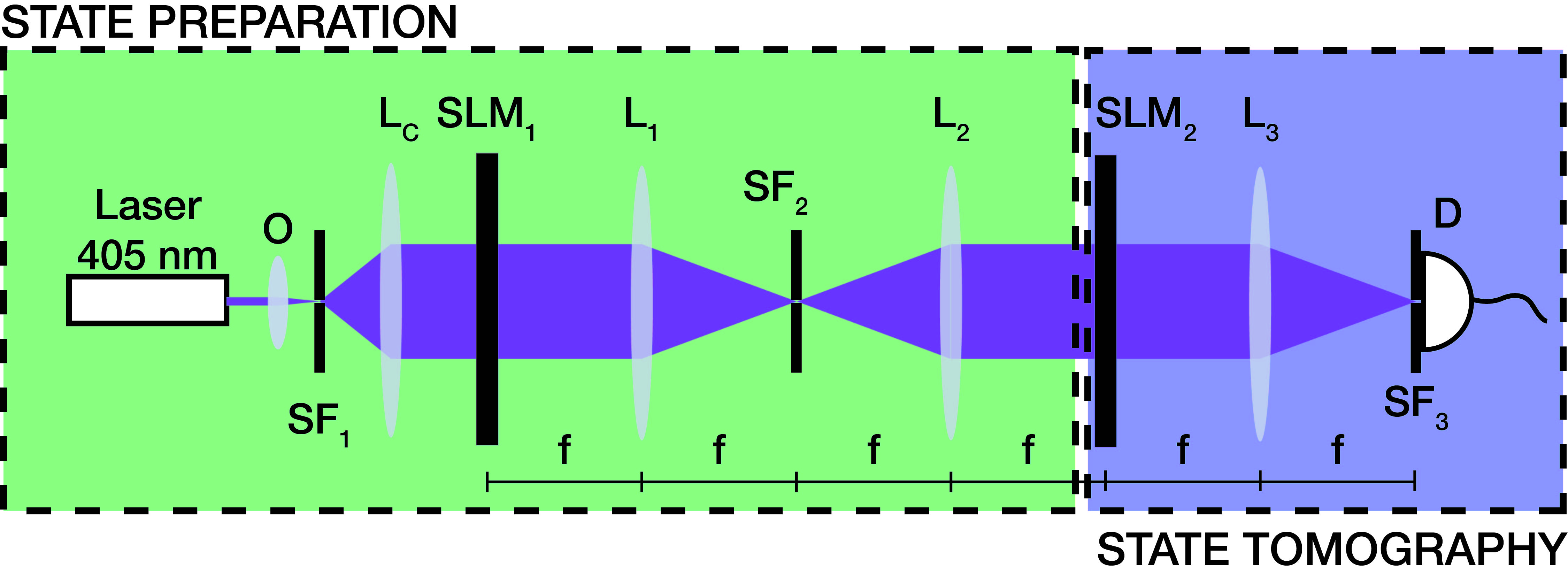}
\caption{Experimental setup. O is a expansor, $\text{SF}_\text{i}$ are spatial filters, $\text{L}_\text{i}$ are lenses, $\text{SLM}_\text{i}$ are spatial light modulators and D is a single pixel detector.} \label{fig:setup}
\end{figure}

The experimental setup used for the generation and reconstruction of
the spatial qudit states is shown, schematically, in
Fig.~\ref{fig:setup}. The first part consists in a $4f$ optical system with a
spatial filter in the Fourier plane.

A $405\text{nm}$ laser diode beam is expanded, filtered and collimated in order to illuminate the spatial light modulator $\text{SLM}_1$ with a planar wave with approximately constant phase and amplitude distribution over the region of interest.
This modulator is used to represent the spatial qudit $|\psi\rangle$ according with the techniques described in~\cite{solis2013,varga2014}. These methods allows us to generate pure spatial qudits with arbitrary complex coefficients by using only one pure phase modulator. The coefficient modulus $\beta_\ell$ (see Sec. \ref{sec:formalism}), is given by the phase modulation of the diffraction gratings displayed on each slit region. The argument $\phi_\ell$ can be defined either by adding a constant phase value~\cite{solis2013} or by means of a lateral displacement of the gratings~\cite{varga2014}. Both methods have a good performance, being
the latter one developed to reduce the effects of the phase fluctuations
present in modern liquid crystal on silicon (LCoS)
displays~\cite{lizana2008}. In particular, the phase modulators used in our experiment, are conformed by a Sony liquid crystal television panel LCTV model LCX012BL in combination with polarizers and wave plates that provide the adequate state of light polarization to reach a phase modulation near $2\pi$~\cite{marquez2001,marquez2008}. As this device is free of phase fluctuations the first codification method was implemented given that it avoids the phase quantization required in the second scheme. The spatial filter $\text{SF}_2$ is used to select the first orders diffracted by the mentioned gratings in such a way that on the back focal plane of lens $\text{L}_2$ is obtained the complex distribution that represents the spatial qudit.

On the same plane (which is also coincident with the front focal plane of $\text{L}_3$) is placed the second modulator $\text{SLM}_2$ on which are represented the reconstruction bases $|\psi_m^{(\alpha)}\rangle$ used to implement the
quantum state tomography process~\cite{lima2011}. These bases are also displayed as slits and its complex amplitudes are codified by following the previously described method. The measurements that allow characterizing the quantum state are performed by means of a single pixel detector placed at the back focal plane of $\text{L}_3$ and a spatial filter $\text{SF}_3$ used to select the center of the interference pattern produced by the slits.

It is worth to mention that the proposed architecture performs the exact Fourier Transform at each stage and avoid the introduction of spurious phases through the propagation process.

\subsection{\label{sec:pure}Generation of pure states}

In order to test the implementation of the encoding method in our optical set-up and optimize the alignment process we started by preparing and reconstructing pure quantum states. 
The generation of pure states is achieved by representing the state $|\psi\rangle$ on the $\text{SLM}_1$, as we explained in Sec.~\ref{sec:exp}. The tomographic process is carried out by means of projective measurements that allow reconstructing the density matrix $\rho$ in Eq.~(\ref{eq:spatial_qudit}). We represent the reconstruction basis $|\psi_m^{(\alpha)}\rangle$ on the $\text{SLM}_2$ and take the number of counts in the center of the Fourier plane as the value of the proyection $p_{\alpha_m}=|\langle\psi_m^{(\alpha)}|\psi\rangle|^2$. We use mutually unbiased basis which require $D(D+1)$ projections and reconstruct the density as~\cite{fernandez2011}

\begin{equation}\label{eq:rho_eq}
\rho=\sum_{\alpha=1}^{D+1}\sum_{m=1}^D p_{\alpha_m} |\psi_m^{(\alpha)}\rangle\langle\psi_m^{(\alpha)}| - I
\end{equation}

To quantify
the quality of the whole experiment we used the fidelity  
$F \equiv \mathrm{Tr}\left(\sqrt{\sqrt{\varrho}\rho\sqrt{\varrho}}\right)$, between the state intended to be prepared,
$\varrho$, and the density matrix of the state actually prepared, $\rho$ \cite{Jozsa1994}. Ideally, it is desirable to have $F =1$. 

\begin{figure}[htbp]
\centering
\includegraphics[width=\linewidth]{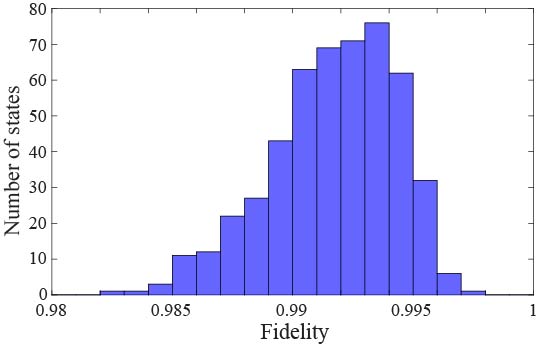}
\caption{Fidelity occurrence for qudits states with $D=11$. There are represented $500$ arbitrary states $|\psi\rangle=\sum_{\ell=0}^{D-1} e^{i\phi_{\ell}}|\ell\rangle$. The mean fidelity is $\overline{F}=0.992$ and the standard deviation is $\sigma=0.003$.}
\label{fig:fid11}
\end{figure}

We have tested the system for different Hilbert space dimensions with excellent results. As an example, the reconstruction results obtained for $D$=11 are shown in Fig.~\ref{fig:fid11}. To this end we have generated $500$ pure states $|\psi\rangle=\sum e^{i\phi_{\ell}}|l\rangle$ with an arbitrary phase $\phi_{\ell}$ uniformly distributed between $0$ and $2\pi$. The mean fidelity is $\overline{F}=0.992$ with standard deviation $\sigma=0.003$. The system proved to be reliable for the generation and reconstruction of pure qudits in different dimensions.

\subsection{\label{sec:mixed}Generation of mixed states}

\begin{figure}[htbp]
\centering
\includegraphics[width=\linewidth]{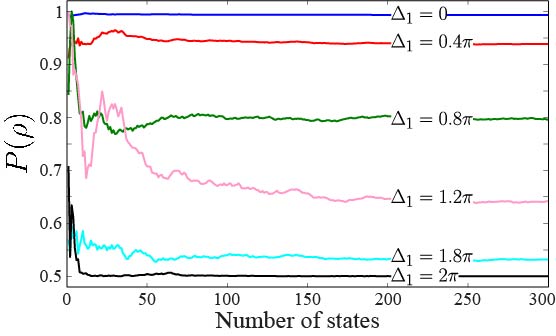}
\caption{Purity evolution for qubits as a function of the number of states composing the statistical mixture, and different probability distribution widths $\Delta_1$. We note that after $250$ pure states were used to generate the mixture, the purity behaviour stabilizes and it achieves its final value.}
\label{fig:evol}
\end{figure}

The mixed states generation is achieved by means a statistical mixture of pure states $|\psi\rangle$. This can be performed by varying the modulus $\vert\beta_{\ell}(t)\vert$ and/or phases $\phi_{\ell}(t)$ of the states represented on $\text{SLM}_1$ while the measurement process is carried on. As previously mentioned, in Sec.~\ref{sec:formalism}, in general, many of the important features of a signal are preserved when only the phase is retained regardless of the amplitude \cite{oppenheim1981}.In order to verify this assertion in our case, we studied, first by numerical simulation, the effect of varying separately these magnitudes. We started by keeping constant the amplitudes and changing the phases with a uniform probability distribution centered on a mean phase value $\phi_{\ell}$, and with a width $\Delta_{\ell}$. The purity of the states are determined by the width $\Delta_{\ell}$ as is stated in Eq.~(\ref{eq:purityD}). The highest incoherence is achieved when $\Delta_{\ell}=2\pi$ for each slit, and narrower widths lead to greater coherence between slits. As pure states are added to the mixture, purity converges to a steady value. As an example, in Fig.~\ref{fig:evol} it is shown the purity evolution as a function of the number of pure states used to generate a mixed state of dimension $D=2$. The evolution is depicted for different width distributions. We can observe that there is a stabilization afterwards $250$ pure states were used to generate the mixture. A similar behaviour was observed for higher dimensions $D$ (see Supplementary Material \cite{sup_mat_1}).

Following the same technique for generating mixed states, we also tested the purity evolution of the states by varying the real amplitudes $\beta_{\ell}$, instead of the phases $\phi_{\ell}$. We have observed that the convergence to a steady purity value is obtained after adding (at least) $500$ pure states in the mixture. Besides, independently of which distribution width $\Delta_{\ell}$ is considered, it is not possible to achieved the lowest purity value. Summarizing, phase variation results the best option in order to generate mixed states.

%--------------------------------------------------

\section{\label{sec:results_mix}Results}

\begin{figure}[htbp]
\centering
\includegraphics[width=\linewidth]{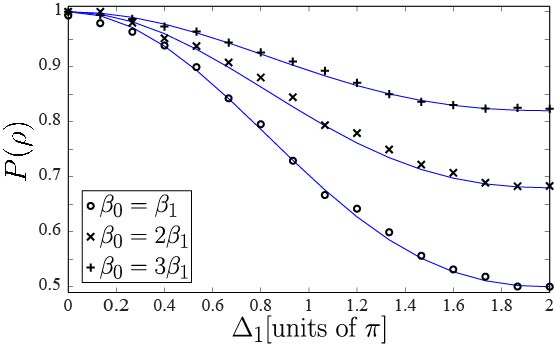}
\caption{Purity of qubits in function of the probability distribution width $\Delta_1$ for different relative amplitudes $\beta_0$ and $\beta_1$. The solid line represents the theoretical values according to Eq.~(\ref{eq:purity2}).}
\label{fig:pur2d}
\end{figure}

In this section are presented and analyzed the results obtained for mixed states ranging from dimension $D=2$ to $D=11$. Let us start with the mixed qubits case. We have generated states with three different relative amplitudes of the two slits ($\beta_0$ and $\beta_1$) and diverse width distributions of the phase variations ($\Delta_1$). The purity of the states, $P(\rho)$, as a function of these magnitudes is shown in Fig.~\ref{fig:pur2d}. In every case the experimental values of purity matches very well with the theoretical behaviour described by equation (\ref{eq:purity2}).

For $\Delta_1=2\pi$ in the case of $\beta_0=\beta_1$, represented with circles, it is possible to reach the lowest purity for qudits, $P(\rho)=\frac{1}{2}$. However, in the cases where $\beta_0=2\beta_1$ (crosses) and $\beta_0=3\beta_1$ (squares) the lowest purity obtained is higher than in the first case. In fact, the value of purity defined by Eq.~ (\ref{eq:purity2}) is function of $\beta_0$, $\beta_1$ and $\Delta_1$ and the lowest purity achievable is when $\beta_0=\beta_1$ and $\Delta_1=2\pi$.

\begin{figure}[htbp]
\centering
\includegraphics[width=\linewidth]{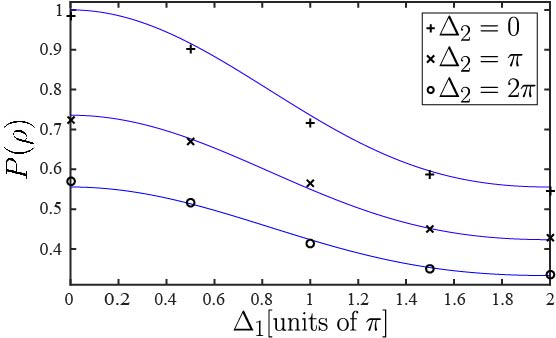}
\caption{Purity of $D=3$ qudits in function of the probability distribution width $\Delta_1$ for different fixed width $\Delta_2$. The relative amplitude are $\beta_0=\beta_1=\beta_2$. The solid line represents the theoretical values according to Eq.~(\ref{eq:purityD}).}
\label{fig:pur3d}
\end{figure}

In case of qutrits ($D=3$), we have generated several mixed states. A particular situation is illustrated in Fig.~\ref{fig:pur3d}. It shows the purity of these states as a function of the probability width $\Delta_1$ of the second slit for different widths $\Delta_2$ fixed on the second slit. The relative amplitudes between slits are $\beta_0=\beta_1=\beta_2$. In the case of $\Delta_2=2\pi$ (circles), the lowest possible purity $P(\rho)=\frac{1}{3}$ is reached for $\Delta_1=2\pi$. Lower values of $\Delta_2$, in this case $\Delta_2=\pi$ (crosses) and $\Delta_2=0$ (plus signs), leads to higher purities. The situation becomes trivial for $\Delta_1=\Delta_2=0$ when it is obtained a pure state ($P(\rho)=1$).

\begin{figure*}[htbp]
\centering
\includegraphics[width=1\linewidth]{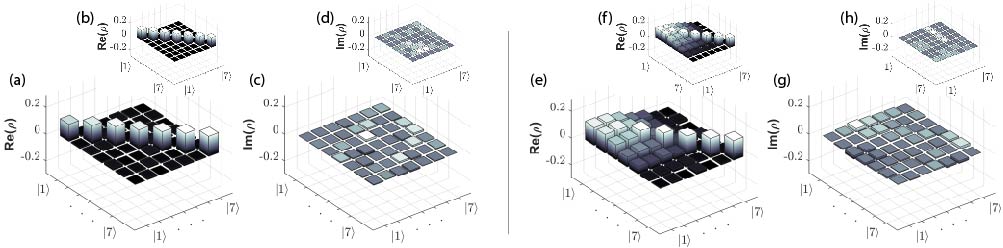}
\caption{Density matrices of $D=7$ mixed states. $(a)$ and $(c)$ are the real and imaginary parts, respectively, when $\Delta_{\ell}=2\pi$ reconstructed after $250$ pure states were used for the mixture. $(b)$ and $(d)$ are the theoretical matrices.  $(e)$ and $(g)$ are the real and imaginary parts, respectively, when $\Delta_{\ell}=\frac{2\pi}{7}\ell$ reconstructed after $250$ iterations. $(f)$ and $(h)$ are the theoretical matrices.} \label{fig:mix7}
\end{figure*}

For $D=7$ we illustrate the case with two different mixed states which density matrices are shown in Fig.~\ref{fig:mix7}. For both states the amplitudes of the slits are equal, i.e., $\beta_0=\beta_1=\dots=\beta_6$. On the left side it is shown the case of lowest coherence between slits, obtained when $\Delta_{\ell}=2\pi$ for every slit. It can be seen that the diagonal elements on the real part (the system populations) are equal and different of zero, while the off diagonal elements (the system coherences) are null. On the right side is shown the density matrix of a state which slit coherences are governed by  probability distribution widths that follow a lineal dependence with the slit label $\ell$, that is $\Delta_{\ell}=\frac{2\pi}{7}\ell$. It can be noted that the system populations remain equal, like in the previous case, but the system coherences decrease as the slit label increases. These examples show that, by means of the proposed method, it is possible to modify the coherence between the slits in an arbitrary way.

\begin{figure*}[htbp]
\centering
\includegraphics[width=1\linewidth]{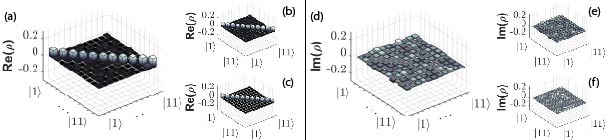}
\caption{Density matrices for $D=11$ when $\Delta_{\ell}=2\pi$. $(a)$ and $(d)$ are the real and imaginary parts, respectively, reconstructed after $250$ pure states were used in the mixture. $(b)$ and $(e)$ are simulated results using the same $250$ states. $(c)$ and $(f)$ are the theoretical matrices.} \label{fig:mix11_2pi}
\end{figure*}

In the case —$D$=11— we present two mixed states with different coherences among slits. Fig.~\ref{fig:mix11_2pi} shows the real (left) and imaginary (right) part of an incoherent state, and thus, with minimal purity. All the relative amplitudes are equal and the phase distribution widths are $\Delta_{\ell}=2\pi$. Fig.~\ref{fig:mix11_2pi}(a) show the reconstructed density matrix by mixing $250$ pure states. Fig.~\ref{fig:mix11_2pi}(b) show simulated results using the same $250$ states and Fig.~\ref{fig:mix11_2pi}(c) are the theoretical density matrix. The fidelity beetween experimental and simulated density matrices is reported as $F=0.9886$. The reported purities are $P(\rho)_{\text{exp}}=0.1201$ and $P(\rho)_{\text{sim}}=0.1177$ for experimental and simulated density matrix, respectively. In this case the lowest purity for a $D=11$ state is $P(\rho)_{\text{theo}}=\frac{1}{11}\sim 0.0909$. We note that the agreement between experimental and simulated results are excellent. The theoretical value correspond to a mixture of infinite pure states and this is the reason for not having reached the maximum incoherence. In the Supplementary Material \cite{sup_mat_1} it is shown a dynamical evolution from the initial pure state to the final mixed state.

\begin{figure*}[htbp]
\centering
\includegraphics[width=1\linewidth]{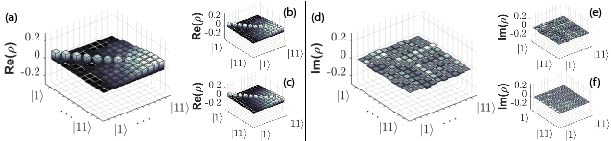}
\caption{Density matrices for $D=11$ when $\Delta_{\ell}=\frac{2\pi}{11}(11-\ell)$. $(a)$ and $(d)$ are the real and imaginary parts, respectively, reconstructed after $250$ pure states used to generate the mixture. $(b)$ and $(e)$ are simulated results using the same $250$ pure states in the mixture. $(c)$ and $(f)$ are the theoretical matrices.}
\label{fig:mix11_lin}
\end{figure*}

Same as in the case $D=7$, for $D=11$ we have generated a mixed state with arbitrary coherences among slits. Fig.~\ref{fig:mix11_lin} shows the real (left) and imaginary (right) parts of this mixed state. In this case the phase distribution width is given by $\Delta_{\ell}=\frac{2\pi}{11}(11-\ell)$. Fig.~\ref{fig:mix11_lin}(a) show the reconstructed density matrix by mixing $250$ pure states. Fig.~\ref{fig:mix11_lin}(b) show simulated results using the same $250$ states and Fig.~\ref{fig:mix11_lin}(c) are the theoretical density matrix. The fidelity between experimental and simulated density matrices is $F=0.9955$. The reported purities are $P(\rho)_{\text{exp}}=0.3152$ and $P(\rho)_{\text{sim}}=0.2870$ for experimental and simulated density matrix, respectively. In this case the lowest purity for a $D=11$ state is $P(\rho)_{\text{theo}}=0.2670$. Additionally a demonstration of the convergence is shown in the Supplementary Material \cite{sup_mat_1}.

%--------------------------------------------------
\section{\label{sec:conclusions}Conclusions}

We have presented a method for the controlled generation of mixed spatial qudits with arbitrary degree of purity. The state generation is achieved by a succession of random pure qudits according to a pre-set probability distribution. We have experimentally showed the viability of the method for qudits from dimension $D$=2 up to $D=11$. The excellent agreement between experimental, simulated and theoretical results demonstrate the feasibility of the method to easily control the coherence between each pair of slits $\ell$ that allow us engineering the state. The method can be extended for the generation of composite systems with controllable degrees of entanglement or mixedness. Besides, it can be used to study the evolution of the system under a specific dynamics since the same technique permit to vary the phases and/or the real amplitude of the slits.

%-----------------------------------------------------

\section*{ACKNOWLEDGMENTS}

This work was supported by UBACyT 20020130100727BA,
CONICET PIP 11220150100475CO, and ANPCYT PICT
2014/2432. J.J.M.V. thanks N.K. and C.F.K. for the heavy heritage.

\bibliography{main}{}
\bibliographystyle{unsrt}

\end{document}